\documentclass{ifacconf}

\usepackage{graphicx}
\usepackage{natbib}        
\usepackage{amsmath}
\usepackage{amssymb}
\usepackage{mathtools}
\usepackage{CJKutf8}
\usepackage{algorithm}
\usepackage{algorithmicx}
\usepackage{algpseudocode}
\usepackage{graphicx}  
\usepackage{epstopdf}
\usepackage{stfloats}
\usepackage{booktabs}
\usepackage{fancyhdr}
\usepackage{lastpage}

\begin{document}
\begin{frontmatter}

\title{A Gauss-Newton-Induced Structure-Exploiting Algorithm for Differentiable Optimal Control\thanksref{footnoteinfo}}


\thanks[footnoteinfo]{This work was supported in part by the National Nature Science Foundation of China under Grant 62573209, in part by the Development and Reform Commission Foundation of Jilin Province under Grant 2023C034-3.}

\author[First,equal]{Yuankun Chen}
\author[First,equal,corr]{Zifei Nie}
\author[First]{Xun Gong}
\author[Second]{Yunfeng Hu}
\author{\quad}
\author[Third]{Hong Chen}

\address[First]{School of Artificial Intelligence, Jilin University, Changchun, China.}
\address[Second]{College of Control Science and Engineering, Jilin University, Changchun, China.}
\address[Third]{College of Electronics and Information Engineering, Tongji University, Shanghai, China.}

\renewcommand{\thefootnote}{\arabic{footnote}}
\thanks[equal]{These authors contributed equally to this work.}
\thanks[corr]{Corresponding author (e-mail: zifei\_nie@jlu.edu.cn).}

\maketitle

\begin{abstract}                
Differentiable optimal control, particularly differentiable nonlinear model predictive control (NMPC), provides a powerful framework that enjoys the complementary benefits of machine learning and control theory. A key enabler of differentiable optimal control is the computation of derivatives of the optimal trajectory with respect to problem parameters, i.e., trajectory derivatives. Previous works compute trajectory derivatives by solving a differential Karush-Kuhn-Tucker (KKT) system, and achieve this efficiently by constructing an equivalent auxiliary system. However, we find that directly exploiting the matrix structures in the differential KKT system yields significant computation speed improvements. Motivated by this insight, we propose \texttt{FastDOC}, which applies a Gauss-Newton approximation of Hessian and takes advantage of the resulting block-sparsity and positive semidefinite properties of the matrices involved. These structural properties enable us to accelerate the computationally expensive matrix factorization steps, resulting in a factor-of-two speedup in theoretical computational complexity, and in a synthetic benchmark \texttt{FastDOC} achieves up to a 180\% time reduction compared to the baseline method. Finally, we validate the method on an imitation learning task for human-like autonomous driving, where the results demonstrate the effectiveness of the proposed \texttt{FastDOC} in practical applications.

\vspace{0.08cm}
Code is available at: \texttt{https://github.com/optiXlab1/FastDOC}
\end{abstract}


\end{frontmatter}

\thispagestyle{fancy}

\section{Introduction}
%

Differentiable optimal control, or more generally differentiable optimization, is emerging as a framework that bridges machine learning and control theory, enabling the deployment of deep learned models in safety-critical applications such as autonomous driving.
From one perspective, optimal control provides inductive biases for learning-based systems, greatly enhancing their safety and interpretability. The research by \cite{zanon2020safe} supports this by using Model Predictive Control (MPC) as a function approximator in Reinforcement Learning (RL), thereby preventing unsafe parameter updates and ensuring safety. In the context of autonomous driving, several works integrate optimal control modules into learning-based frameworks, such as combining perception networks with MPC (\cite{huang2023differentiable}) or Control Barrier Functions (\cite{xiao2023barriernet}) to enforce safety constraints and improve decision-making interpretability.
From another perspective, learning-based approaches enable optimal control to adjust its strategies based on guidance--such as demonstration data or reward signals--thereby enabling more adaptive control policies. In terms of strategy adjustment for imitating demonstrations, \cite{Huang2023} focuses on personalized lane-change control, leveraging human behavioral demonstrations to model driver preferences. Similarly, \cite{Nie2024} infers control objectives from human cruise driving data to adapt the system to diverse driving styles. In terms of adjusting strategies based on reward signals, \cite{adabag2025DMPC} integrates reinforcement learning to optimize strategies based on reward guidance, thereby increasing the success rate of automated drifting.

Although differentiable optimal control has become increasingly mature, its adoption in deep learning pipelines is still hindered by its substantial computational cost. Compared with conventional lightweight neural network layers--such as linear layers, convolutional layers, and attention-based layers--the \textbf{forward pass} and \textbf{backward pass} of differentiable optimal control remain considerably more expensive. To articulate its computation burden issue, consider a parametric nonlinear optimal control problem (OCP) with finite horizon $N$:
\begin{equation}
	\label{eq:OCP}
	\tag*{$(\mathcal{P})$}
	\begin{aligned}
		\mathop{\text{minimize}} \limits_{x, u} \;\;
		&J(x,u;\theta)=\sum_{k=0}^{N-1} \ell_k(x_k,u_k; \theta) + \ell_N(x_N; \theta) \\
		\text{subject to} \,\,
		& x_0 = x_{\text{init}},\\
		& x_{k+1} = f_k(x_k,u_k; \theta),\;k \in\mathbb{N}_{\left[0,N-1\right]}, \\
		& g_k(x_k,u_k; \theta) \le 0,\; k \in\mathbb{N}_{\left[0,N-1\right]}, \\
		& h_k(x_k,u_k; \theta) = 0,\;k \in\mathbb{N}_{\left[0,N-1\right]}, \\
		& g_N(x_N; \theta) \le 0,\\
		& h_N(x_N; \theta) = 0,
	\end{aligned}
\end{equation}
where the function $f_k : \mathbb{R}^{n_x} \times \mathbb{R}^{n_u} \to \mathbb{R}^{n_x}$ models the discrete-time, nonlinear dynamics of the system, with initial state $x_{\text{init}}$. The functions $h_k : \mathbb{R}^{n_x} \times \mathbb{R}^{n_u} \to \mathbb{R}^{n_h}$ and $g_k : \mathbb{R}^{n_x} \times \mathbb{R}^{n_u} \to \mathbb{R}^{n_g}$ represent the stage equality and inequality constraints, $h_N : \mathbb{R}^{n_x} \to \mathbb{R}^{n_h}$ and $g_N : \mathbb{R}^{n_x} \to \mathbb{R}^{n_g}$ represent the terminal equality and inequality constraints, and $\ell_k : \mathbb{R}^{n_x} \times \mathbb{R}^{n_y} \to \mathbb{R}$ and $\ell_N : \mathbb{R}^{n_x} \to \mathbb{R}$ define the stage costs and the terminal cost respectively. 

\begin{figure}[t]
	\centering
	\includegraphics[width=\linewidth, trim=8cm 3cm 8cm 1cm, clip]{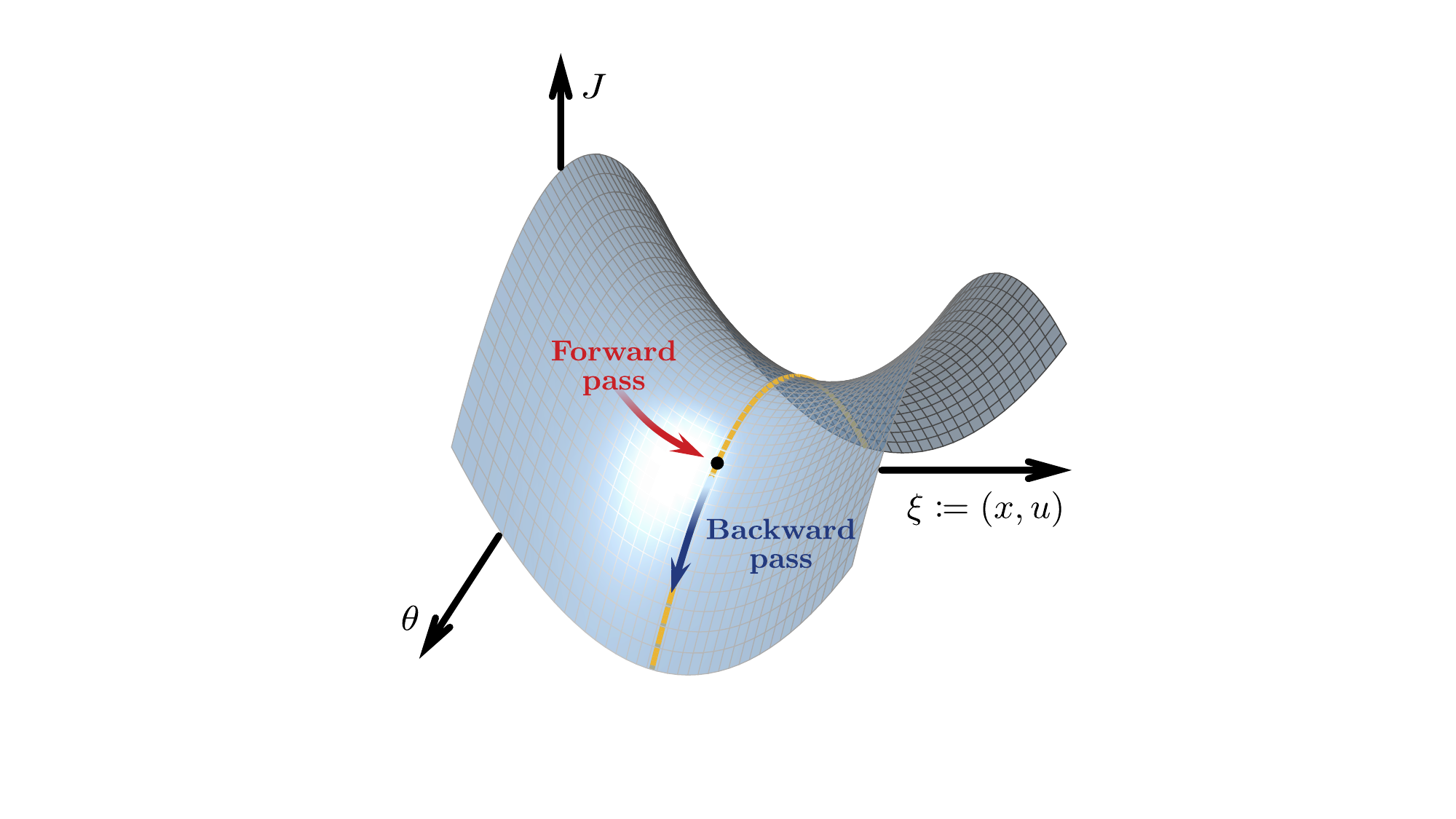}
	\caption{The forward and backward pass in differentiable optimal control.}
	\label{fig:fbd}
\end{figure}

As shown in Fig.~\ref{fig:fbd}, the \textbf{forward pass} refers to solving the optimization problem in \ref{eq:OCP} to obtain the optimal state and control trajectory \((x^{\star}, u^{\star})\) with minimum cost, and the \textbf{backward pass} refers to evaluating the sensitivity of the optimal solution with respect to the problem parameters \(\theta\). The resulting sensitivities, \(\frac{\partial x^{\star}}{\partial \theta}\) and \(\frac{\partial u^{\star}}{\partial \theta}\), are referred to as trajectory derivatives. 
For nonlinear problem \ref{eq:OCP}, the \textbf{forward pass} is often computationally demanding, mainly due to the difficulty of achieving convergence under nonlinear costs and constraints, but many mature and widely used techniques have been developed to address this issue, such as CasADi (\cite{casadi}) and \texttt{acados} (\cite{acados}). The \textbf{backward pass} is also computationally expensive, as it requires solving a high-dimensional linear system derived from the optimality conditions of the problem \ref{eq:OCP}, and our work therefore focuses on improving the computational efficiency of the \textbf{backward pass}.

For the \textbf{backward pass}, most differentiable solvers--the algorithms computing the trajectory derivatives--focus on the differential KKT system, which is derived by applying the Implicit Function Theorem (IFT) to the KKT conditions of \ref{eq:OCP} (\cite{amos2017optnet, IMPC}), because solving the differential KKT system yields the trajectory derivatives of interest. To accelerate the evaluation of the differential KKT system, one line of research involves constructing an equivalent auxiliary system. From the perspectives of optimality conditions (\cite{NEURIPS2024_BPQP}) and Pontryagin’s maximum principle (\cite{PDP,SafePDP}), prior works formulate auxiliary systems whose solutions are equivalent to the solutions of the differential KKT system, thereby making the computation of trajectory derivatives more efficient. The work by \cite{frey2025DNMPC} further integrates the forward pass with these auxiliary systems, avoiding the construction of certain variables and thereby providing additional speedup. However, the solutions obtained from such auxiliary systems may not always coincide with the true trajectory derivatives, due to numerical errors introduced by active set changes, which can lead to unreliable results.

Another line of work solves the differential KKT system directly by exploiting its problem special structure. The research by \cite{DDN} derives an explicit formulation for trajectory derivatives using a Schur-complement-based decomposition, while \cite{IDOC} further accelerates the computation by exploiting the block-sparse structure of the matrices in the explicit formulation. However, these methods still do not fully leverage the structural properties of the matrix discussed above.

Based on the above considerations, we propose \textbf{FastDOC}, a Gauss-Newton-Induced Structure-Exploiting Algorithm, to accelerate the computation of trajectory derivatives. Under the Gauss-Newton approximation, the matrices in the differential KKT system, which exhibit block-sparsity, also features the positive semidefinite (PSD) property. Crucially, by exploiting these properties, the most computationally intensive step in computing trajectory derivatives--LU factorization of large-scale linear systems--is replaced by a series of small-scale and efficient Cholesky factorizations, resulting in a theoretical reduction in time complexity.

Our contributions can thus be summarized as follows:
\begin{itemize}
	\item \texttt{FastDOC} is proposed to efficiently compute trajectory derivatives. Under a Gauss-Newton approximation, \texttt{FastDOC} exploits both the block-sparsity and the PSD properties of the matrices in the differential KKT system, enabling the acceleration of the time-consuming matrix factorization and yielding a factor-of-two theoretical speedup. On a comprehensive synthetic benchmark, \texttt{FastDOC} achieves an average 180\% speedup over the baseline.
	
	\item \texttt{FastDOC} is further validated in imitation learning tasks for autonomous driving trajectory tracking, where human driving demonstrations are collected through a driving-simulator platform, and the model is designed to incorporate personalized driver characteristics. The experimental results demonstrate the effectiveness of \texttt{FastDOC}.
\end{itemize}

\section{Gauss-Newton-Induced Structure-Exploiting Algorithm}
In this section, we first cast the optimization problem \ref{eq:OCP} into a compact form, then derive an explicit identity for the trajectory derivatives by leveraging first-order optimality conditions and the implicit function theorem, followed by the introduction of our acceleration algorithms and underlying principles. Finally, we compare our proposed techniques with several representative differential solvers to demonstrate their effectiveness.

\subsection{Explicit Formulation of Trajectory Derivatives}
To begin with, we define the notation of derivatives. 
For \( f(x): \mathbb{R}^n \to \mathbb{R}^m \), the derivative of the vector-valued function with respect to \( x \) is denoted as \( \nabla_x f \in \mathbb{R}^{m \times n} \), where \( (\nabla_x f)_{i,j} = \frac{\partial f_i}{\partial x_j} \).
For \( f(x, y): \mathbb{R}^n \times \mathbb{R}^m \to \mathbb{R} \), the second-order  derivative with respect to \( x \) and \( y \) is written as \( \nabla_{x,y}^2 f \), which is equivalent to \( \nabla_y (\nabla_x f)^\top \).

To better expose the time-induced structure inherent in the nonlinear OCP, we recast the
problem \ref{eq:OCP} into a compact form as follows
\begin{equation}
	\label{eq:COCP}
	\begin{aligned}
		\mathop{\text{minimize}}_{\xi}\quad 
		& J(\xi;\theta)
		= \sum_{k=0}^{N-1} \ell_k(\xi_k;\theta) + \ell_N(\xi_N;\theta) \\
		\text{subject to}\quad 
		& c(\xi;\theta)=0,
	\end{aligned}
\end{equation}
where $\xi \coloneqq \left( x,u \right)$ denotes the stacked decision variable containing states and controls, and  
$c$ consumes the stacked constraints. The stacked decision variable
$
\xi = (\xi_0,\,\xi_1,\,\ldots,\,\xi_N)
$
with
\begin{equation}
	\label{eq:trajectory-stacking}
	\xi_k =
	\begin{cases}
		(x_k,\,u_k), & k = 0,\ldots,N-1, \\[4pt]
		x_N,       & k = N,
	\end{cases}
\end{equation}
and the stacked constraints
$
c = (c_{-1},\,c_0,\,\ldots,\,c_N)
$
with
\begin{equation}
	\label{eq:constraint-stacking}
	\resizebox{0.9\hsize}{!}{$
			c_k =
		\begin{cases}
			x_0 - x_{\mathrm{init}}, & k = -1, \\[4pt]
			\bigl(\tilde g_k(\xi_k;\theta), h_k(\xi_k;\theta), f_k(\cdot)\,\bigr), 
			& k = 0,\ldots,N-1, \\[4pt]
			\bigl(\tilde g_N(x_N;\theta), h_N(x_N;\theta)\bigr), & k = N,
		\end{cases}
		$}
\end{equation}
the $\tilde g_k$ denotes the numerically identified active inequality constraints at time $k$, selected using a threshold~$\varepsilon$.

To facilitate the computation of $\nabla _{\theta}$ in \eqref{eq:COCP}, we first introduce the Lagrange multipliers
$\lambda \in \mathbb{R}^{n_{r}}$ and construct the corresponding Lagrangian as
\begin{equation}
\mathcal{L}(\xi,\lambda;\theta)\;=\; J(\xi;\theta)+\lambda^\top c(\xi;\theta).
\end{equation}

The Karush-Kuhn-Tucker (KKT) conditions enforcing stationarity and primal feasibility are accordingly take the following form 
\begin{equation}
	\label{eq:KKT}
	\begin{bmatrix}
		\nabla_\xi \mathcal{L}(\xi, \lambda; \theta)\\[3pt]
		\nabla_\lambda \mathcal{L}(\xi, \lambda; \theta)
	\end{bmatrix}
	=
	\begin{bmatrix}
		\nabla_\xi J(\xi;\theta)+\nabla_\xi c(\xi;\theta)^\top\lambda\\[3pt]
		c(\xi;\theta)
	\end{bmatrix}
	= 0 .
\end{equation}

If there exists solution $(\xi^\star,\lambda^\star)$ that satisfies \eqref{eq:KKT}, it is called a KKT point. And such a KKT point is a local minimum of \eqref{eq:COCP} if the following assumption holds:

\textit{\textbf{Assumption 1.}}
\label{ass:reg}
The function $J$ and $r$ are twice continuously differentiable with respect to $\xi$
and once continuously differentiable with respect to $\theta$.
Moreover, let $(\xi^\star,\lambda^\star)$ be a KKT point  that satisfies the linear independence constraint qualification (LICQ) and the second-order sufficient conditions of optimality (SOSC).

Recall that our goal is to obtain $\nabla_{\theta}\xi^\star$ and $\nabla_{\theta}\lambda^\star$, 
To this end, we are motivated to differentiate the KKT system in \eqref{eq:KKT} by leveraging the IFT. This leads to the differential KKT linear system as follows

\begin{equation}
	\label{eq:dKKT}
	\begin{bmatrix}
		H & A^\top\\[2pt]
		A & 0
	\end{bmatrix}
	\begin{bmatrix}
		\nabla_{\theta}\xi^\star\\[4pt]
		-\,\nabla_{\theta}\lambda^\star
	\end{bmatrix}
	=
	-\begin{bmatrix}
		B\\[2pt]
		C
	\end{bmatrix}.
\end{equation}
where
\[
\begin{array}{cc}
	\displaystyle
	H =
	\begin{bmatrix}
		H_0 & & & \\
		& H_1 & & \\
		& & \ddots & \\
		& & & H_N
	\end{bmatrix},
	&
	\displaystyle
	B =
	\begin{bmatrix}
		B_0 \\[2pt]
		B_1 \\[2pt]
		\vdots \\[2pt]
		B_N
	\end{bmatrix},
	\\[10pt]
	\displaystyle
	A =
	\begin{bmatrix}
		A_{-1,0} & & & \\
		A_{0,0} & A_{0,1} & & \\
		& \ddots & \ddots & \\
		& & A_{N-1,N-1} & A_{N-1,N} \\
		& & & A_{N,N}
	\end{bmatrix},
	&
	\displaystyle
	C =
	\begin{bmatrix}
		C_0 \\[2pt]
		C_1 \\[2pt]
		\vdots \\[2pt]
		C_N
	\end{bmatrix}.
\end{array}
\]
with
$H_k = \nabla_{\xi_k\xi_k}^2 \mathcal{L}$, $A_{i,j} = \nabla_{\xi_j} c_i$, $B_k = \nabla_{\xi_k\theta} \mathcal{L}$, and $C_k = \nabla_{\theta} c_k$.

\textbf{Proposition 1.}
	Under assumption 1, the analytical trajectory derivatives associated with the nonlinear OCP in \eqref{eq:COCP} can be expressed as 
	\begin{equation}
		\label{eq:explicit}
		\resizebox{0.9\hsize}{!}{$
			\nabla_{\theta}\xi^\star
			= H^{-1}A^\top\bigl(AH^{-1}A^\top\bigr)^{-1}
			\bigl(AH^{-1}B - C\bigr)-H^{-1}B.
			$}
	\end{equation}

\begin{pf}
	Such an analytical result is obtained by solving the linear system in \eqref{eq:dKKT}:
	the derivation begins with the Schur complement of the matrix H, which is given by
	\begin{equation}
		\label{eq:SandGamma}
		S := A H^{-1} A^\top, 
		\qquad 
		\Gamma := A H^{-1} B - C.
	\end{equation}
	With these definitions, the derivatives \(\nabla_{\theta}\lambda^\star\) and 
	\(\nabla_{\theta}\xi^\star\) can be obtained sequentially.
	The first relation follows from eliminating $\nabla_{\theta} \xi^{\star}$:
	\begin{equation}
		\label{eq:schursub-1}
		S\,\nabla_{\theta}\lambda^\star = \Gamma,
	\end{equation}
	Having solved for $\nabla_{\theta} \xi^{\star}$, the remaining derivative is recovered from the back-substitution: 
	\begin{equation}
		\label{eq:schursub-2}
		\nabla_{\theta}\xi^\star = H^{-1}(A^\top \nabla_{\theta}\lambda^\star - B).
	\end{equation}
	Substituting the result of \eqref{eq:schursub-1} into \eqref{eq:schursub-2} explicitly yields the analytical expression 
	in \eqref{eq:explicit}.
	\hfill $\square$
\end{pf}

\textit{\textbf{Remark 1.}} (Block sparsity)\label{rem:block-sparsity}
	Both $H$ and $A$ exhibit a block-sparse structure induced by the temporal dependencies in $r$ and $\xi$, together with the time-additive nature of the stage cost. In particular, each block $r_k$ depends only on $\xi_k$ and, when applicable, $\xi_{k+1}$, whereas the stage cost $\ell_k$ depends solely on $\xi_k$. These locality properties restrict the nonzero patterns of $H$ and $A$ to a small set of block diagonals, thereby yielding block-sparse matrices. This sparsity structure not only enables memory-efficient storage but also permits effective blockwise parallel computation of matrix products across all time steps, including $H^{-1}A^\top$, $A H^{-1}A^\top$, $H^{-1}B$, and $A H^{-1}B$. Further implementation details about leveraging block-sparse structure are provided in research work by~\cite{IDOC}.

\textit{\textbf{Remark 2.}} (Runtime cost)\label{rem:runtime}
	To assess the computational cost associated with \eqref{eq:explicit}, the overall procedure can be decomposed into four constituent steps:
	
	\textbf{Step 1:} formation of $H^{-1}$;
	
	\textbf{Step 2:} construction of $S$ and $\Gamma$ in \eqref{eq:SandGamma};
	
	\textbf{Step 3:} solution $S\,\nabla_{\theta}\lambda^\star = \Gamma$ in \eqref{eq:schursub-1} to obtain $\nabla_{\theta}\lambda^\star$;
	
	\textbf{Step 4:} back-substitution to compute $\nabla_{\theta}\xi^\star$ in \eqref{eq:schursub-2}.
	
	As reported in Table~1, Step~1 and Step~3 dominate the total runtime, together accounting for more than $90\%$ of the overall cost. This is primarily due to the matrix factorizations required in these stages. In contrast, Step~2 and Step~4 leverage the block-sparse structure of $H$ and $A$ and therefore incur only a comparatively minor computational burden. This pronounced disparity in computational cost motivates the development of accelerated schemes for Step~1 and Step~3, which are introduced in the following sections.

\begin{table}[h]
	\centering
	\caption{Runtime cost.\protect\footnotemark}
	\label{tab:runtime_breakdown}
	\begin{tabular}{lr}
		\hline
		Component & Time (\%) \\
		\hline
		Step 1  & 56.75 \\
		Step 2  &  5.72 \\
		Step 3  & 33.56 \\
		Step 4  &  3.98 \\
		\hline
	\end{tabular}
\end{table}
\footnotetext{Runtime performance is evaluated on a synthetic benchmark (Sec.~\ref{sec:eval}) using multiple metrics, with results averaged across all measurements.}

\subsection{Gauss–Newton–Induced Structure for Efficient Cholesky Factorization}
To reduce the computational cost of evaluating \eqref{eq:explicit}, we concentrate on accelerating the matrix factorizations that constitute the dominant runtime bottleneck. Leveraging the Gauss-Newton-induced structure, the Hessian admits a positive semidefinite (PSD) approximation whose structure permits the use of efficient Cholesky factorizations, thereby substantially lowering the associated computational burden. The validity of the Gauss-Newton approximation relies on the following assumption.

\textit{\textbf{Assumption 2.}}
\label{ass:GN}
	The stage cost can be admitted to the composition
	$
	\ell_k(\xi;\theta)=\tfrac12\|\phi(\xi;\theta)\|^2,
	$
	where $\phi$ is smooth.
	Moreover, $\phi(\xi;\theta)$ is close to zero (small residual regime) around the neighborhood of optimal solution $\xi^\star$.

Then the approximate Hessian is formed as:
\begin{equation}
	\label{eq:GNHessian}
	H_\ast \;=\;
	\nabla_{\xi} \phi(\xi;\theta)^{\top}\,
	\nabla_{\xi} \phi(\xi;\theta).
\end{equation}
In the following sections, all instances of the symbol $H$ refer to the approximate Hessian $H_\ast$. Since \( H \) only requires the computation of first-order derivatives, it reduces the time cost of matrix construction. More importantly, \( H \) is PSD by construction. Leveraging this property, we accelerate the following two time-consuming steps:

\textbf{Computation of $H^{-1}$.} 
Rather than evaluating $H^{-1}$ explicitly, we compute the required inverse operation by solving the linear system $HX = I$, using a matrix factorization followed by forward and backward substitutions. The resulting procedure is summarized in \textbf{Algorithm~1}, which consists of a factorization phase and subsequent forward- and backward-substitution phases. Exploiting the block-sparse structure of $H$, the factorization is carried out independently on each block, instead of on the full $T(n+m)$-dimensional system. Moreover, thanks to the PSD structure of $H$, we employ a Cholesky factorization in place of a general-purpose LU factorization. In terms of floating-point operations, this reduces the leading constant in the cubic complexity from $\mathcal{O}\!\left(\tfrac{2}{3}n^{3}\right)$ to $\mathcal{O}\!\left(\tfrac{1}{3}n^{3}\right)$, corresponding to an approximate factor-of-two improvement.

\begin{algorithm}[t]
	\caption{Computation of $H^{-1}$}
	\label{alg:chol_block_inverse}
	\begin{algorithmic}[1]
		\Statex \textbf{Input:} Block-diagonal symmetric positive semidefinite matrix 
		$H = \mathrm{diag}(H_0,\dots,H_N)$, 
		where each block $H_k \in \mathbb{R}^{n_\xi\times n_\xi}$.
		
		\For{$k = 0,\dots,N$}
		
		\State \scalebox{0.95}{\textbf{Cholesky factorization:} Solve $H_k = L_k L_k^\top$ for $L_k$.}
		\Statex \hspace{0.42cm} For $i = 1,\dots,n_\xi$:
		\Statex \hspace{0.42cm}\qquad $(L_k)_{ii} \gets 
		\sqrt{\, (H_k)_{ii} - \sum_{m=1}^{i-1} (L_k)_{im}^2 \, }$
		\Statex \hspace{0.42cm} For $j = 1,\dots,i-1$:
		\Statex \hspace{0.42cm}\qquad $(L_k)_{ij} \gets 
		\dfrac{ (H_k)_{ij} - \sum_{m=1}^{j-1} (L_k)_{im}(L_k)_{jm} }
		{ (L_k)_{jj} }$
		
		\State \scalebox{0.95}{\textbf{Forward substitution:} Solve $L_k Y_k = I$ for $Y_k$.}
		\Statex \hspace{0.42cm} For each column $e_r$ of $I$:
		\Statex \hspace{0.42cm}\qquad
		$(Y_k)_{ri} \gets \dfrac{1}{(L_k)_{ii}} 
		\left( (e_r)_i - \sum_{m=1}^{i-1} (L_k)_{im} (Y_k)_{rm} \right)$
		
		\State \scalebox{0.95}{\textbf{Backward substitution:} Solve $L_k^\top X_k = Y_k$ for $X_k$.}
		\Statex \hspace{0.42cm} For each column $(Y_k)_r$:
		\Statex \hspace{0.42cm}\qquad
		$(X_k)_{ri} \gets \dfrac{1}{(L_k)_{ii}}
		\left( (Y_k)_{ri} - \sum_{m=i+1}^{n_\xi} (L_k)_{mi} (X_k)_{rm} \right)$
		
		\State \textbf{Store} $X_k$ as the $k$-th diagonal block of $X$.
		
		\EndFor
		
		\State \textbf{Return:} $X = H^{-1}$.
		
	\end{algorithmic}
\end{algorithm}

\textbf{Computation of $(A H^{-1}A^{\top})^{-1}(A H^{-1}B - C)$.}
Using the notation from \eqref{eq:SandGamma}, this is equivalent to solving the linear system
$S X = \Gamma$, where $S$ features a block tridiagonal structure as follows:
\begin{equation}
	S =
	\begin{bmatrix}
		Q_0 & U_0 & & \\
		L_0 & Q_1 & U_1 & \\
		& \ddots & \ddots & \ddots \\
		& & L_{N-2} & Q_{N-1} & U_{N-1} \\
		& & & L_{N-1} & Q_N
	\end{bmatrix},
\end{equation}
with the diagonal blocks $Q_k$, lower diagonal blocks $L_k$, and upper diagonal blocks $U_k$ are given by
\begin{align*}
	Q_k &= 
	\begin{cases}
		A_{0,0} H_0^{-1} A_{0,0}^\top, 
		& k = 0, \\[4pt]
		A_{k-1,k} H_k^{-1} A_{k-1,k}^\top 
		+ A_{k,k} H_k^{-1} A_{k,k}^\top,
		& k = 1,...,N,
	\end{cases} \\[6pt]
	L_k &= A_{k,k} H_k^{-1} A_{k-1,k}^\top, \qquad k = 0,...,N , \\[4pt]
	U_k &= A_{k-1,k} H_k^{-1} A_{k,k}^\top. \qquad k = 0,...,N .
\end{align*}

To solve the block tridiagonal linear system $SX=\Gamma$ we employ the recursive block tridiagonal algorithm summarized in \textbf{Algorithm~2}. The method consists of a forward sweep, which constructs the Schur-complement blocks $\widetilde{Q}_k$ together with the modified right-hand sides $\widetilde{\Gamma}_k$, followed by a backward substitution that recovers the solution blocks $X_k$, from $t=N$ down to $t=0$. Throughout both phases, the linear systems that must be solved all have coefficient matrices $\widetilde{Q}_k$. In the next step, we show that each $\widetilde{Q}_k$ is PSD, which implies that all solves involving $\widetilde{Q}_k$ can be carried out efficiently using Cholesky factorizations, allowing us to apply exactly the same Cholesky-based procedure as in \textbf{Algorithm~1} to accelerate the computation.
\begin{algorithm}[t]
	\caption{Computation of $S^{-1}\Gamma$}
	\label{alg:block_thomas}
	\begin{algorithmic}[1]
		
		\Statex \textbf{Input:} Block tridiagonal matrix $S$ with blocks 
		$Q_k, L_k, U_k$ for $k=0,\dots,N$, 
		and block vector $\Gamma = (\Gamma_0,\dots,\Gamma_N)$.
		
		\State Initialize $\widetilde{Q}_0 \gets Q_0$, 
		$\widetilde{\Gamma}_0 \gets \Gamma_0$.
		
		\For{$k = 1,\dots,N$}
		\State Solve $\widetilde{Q}_{k-1} \widetilde{Q}_{k-1}^{-1} = I$ for $\widetilde{Q}_{k-1}^{-1}$.
		\State $G_{k-1} \gets \widetilde{Q}_{k-1}^{-1} L_{k-1}$.
		\State $\widetilde{Q}_k \gets Q_k - G_{k-1} U_{k-1}$.
		\State $\widetilde{\Gamma}_k \gets \Gamma_k - G_{k-1} \widetilde{\Gamma}_{k-1}$.
		\EndFor
		
		\State Solve $\widetilde{Q}_N X_N = \widetilde{\Gamma}_N$ for $X_N$.
		\For{$t = N-1,\dots,0$}
		\State Solve $\widetilde{Q}_k X_k = 
		\widetilde{\Gamma}_k - U_k X_{k+1}$ for $X_k$.
		\EndFor
		
		\State Assemble the full solution $X \gets (X_0,\dots,X_N)$.
		
		\State \textbf{Return:} Solution $X = S^{-1}\Gamma$.
		
	\end{algorithmic}
\end{algorithm}

\textbf{Proposition 2.}\label{prop:PSD}
Given that the Hessian matrix \( H \succeq 0 \), then the updated diagonal blocks in \textbf{Algorithm 2} satisfy \( \widetilde{Q}_k \) \( \succeq 0 \).

\begin{pf}
	Consider the matrix \( S := A H^{-1} A^\top \). Since \( H^{-1} \succeq 0 \) on the range of \( H \), it follows that \( S \succeq 0 \). Specifically, for any vector \( v \), we have:
	\begin{equation}
		v^\top S v = v^\top A H^{-1} A^\top v = (A^\top v)^\top H^{-1} (A^\top v),
	\end{equation}
	which implies \( v^\top S v \geq 0 \), and hence \( S \succeq 0 \). By the same reasoning, each diagonal block \( Q_k \) of \( S \) is also PSD.
	
	Next, consider the block matrix \( M^{(0)} \) at step \( k = 0 \):
	\begin{equation}
		M^{(0)} = \begin{bmatrix} 
			\widetilde{Q}_0 & U_0 \\
			L_0 & Q_1
		\end{bmatrix}.
	\end{equation}
	Since \( S \succeq 0 \), it follows that the principal submatrix \( M^{(0)} \succeq 0 \), and \( \widetilde{Q}_0 = Q_0 \succeq 0 \). The forward elimination step updates the block \( \widetilde{Q}_1 \) as follows:
	\begin{equation}
		\widetilde{Q}_1 = Q_1 - L_0 \widetilde{Q}_0^{-1} U_0.
	\end{equation}
	By the property of Schur complements, we conclude that \( \widetilde{Q}_1 \succeq 0 \).
	Since forward elimination is a congruence transformation, it preserves the PSD of the matrix. Therefore, the updated matrix \( S_{\text{new}} \) remains PSD.
	The updated matrix \( S_{\text{new}} \) takes the following block form:
	\begin{equation}
		S_{\text{new}} = \begin{bmatrix} 
			\widetilde{Q}_0 & & & & \\
			& \widetilde{Q}_1 & U_1 & \\
			& L_1 & Q_2 & \ddots \\
			& & \ddots & \ddots
		\end{bmatrix}.
	\end{equation}
	Thus, \( M^{(1)} \succeq 0 \). By applying the same property of Schur complements, it follows that \( \widetilde{Q}_2 \succeq 0 \). Through this iterative process, all updated diagonal blocks \( \widetilde{Q}_k \succeq 0 \).
	\hfill $\square$
\end{pf}

Therefore, with the Gauss--Newton approximation, all heavy LU factorizations in the 
derivative computation can be replaced by more efficient Cholesky 
factorizations. Based on this simplification, our complete procedure for 
computing trajectory derivatives is presented in \textbf{Algorithm~3}.

\begin{algorithm}[t]
	\caption{FastDOC Backward Pass}
	\label{alg:fastdoc_backward}
	\begin{algorithmic}[1]
		\setlength{\itemsep}{2pt}  
		\Statex \textbf{Input:} Optimal solution $\xi^{*}$ with Lagrange multipliers $\lambda^{*}$ in the forward pass.
		\State Compute approximated Hessian $H$ using \eqref{eq:GNHessian}.
		\State Solve $HX=I$ to obtain $H^{-1}$ by Algorithm 1.
		\State Compute $S$ and $\Gamma$ using \eqref{eq:SandGamma}.
		\State Solve $ SX = \Gamma $ to obtain $\nabla_\theta \lambda^{*}$ by Algorithm 2.
		\State Compute $\nabla_\theta \xi^{*}$ using \eqref{eq:schursub-2}.
		\State \textbf{Return:} Trajectory derivatives $\nabla_\theta \xi^{*}$.
	\end{algorithmic}
\end{algorithm}

\subsection{Comparative Performance Evaluation Against Existing Differentiable Solvers}
\label{sec:eval}

\vspace{-0.6em}
This section presents a comparative evaluation of the proposed algorithm against several baselines for the computation required in \eqref{eq:explicit}. Our goal is to assess both the acceleration achieved by our method and its scaling behavior with respect to key problem parameters. To provide a controlled and reproducible setting, we construct synthetic test instances that allow systematic variation of the problem dimensions and structural characteristics. The resulting benchmark enables us to evaluate computational efficiency and numerical stability in a broad range of scenarios.

\noindent\textbf{Synthetic Benchmark.} 
To construct the synthetic benchmark, we randomly generate the block components required to form the structured matrices $\{H, A, B, C\}$ in~\eqref{eq:dKKT}. The matrix $H$ is obtained as a positive semidefinite matrix by sampling a random matrix and symmetrizing it, after which its spectrum is scaled to yield a condition number of $10^3$. The matrix $A$ is generated as a full-rank block matrix, while the remaining blocks $B$ and $C$ are sampled in a manner consistent with the problem dimensions.

\noindent\textbf{Baselines.}
We compare our method with two representative differentiable solvers: \texttt{SafePDP} (\cite{SafePDP}), which computes trajectory derivatives via an auxiliary system, and \texttt{IDOC} (\cite{IDOC}), which evaluates them explicitly while considering the blocked structure of problem formulation. For all methods, the forward OCP is solved using \texttt{IPOPT} through CasADi (\cite{casadi}), and the associated Lagrange multipliers $\lambda$ are obtained directly from the solver outputs.

\noindent\textbf{Experiment Setup.}
We systematically vary:
\begin{itemize}
	\vspace{-0.4em}
	\item the horizon length \(N\),
	\item the state variable dimension \(n\) (with control variable dimension \(m = n/4\)),
	\item the number of parameters \(d\).
\end{itemize}
\vspace{-0.4em}
In each configuration, only one variable is varied while all others are fixed at their base values.
Each setting is evaluated over 20 trials, and the mean and standard deviation of the runtime are reported. All experiments are run on a single thread of an Intel(R) Core(TM) i5-12600KF CPU@3.7 GHz with 32GB RAM.

\begin{figure*}[!t]
	\begin{center}
		\includegraphics[width=0.9\textwidth,trim=0cm 0.8cm 0cm 0cm, clip]{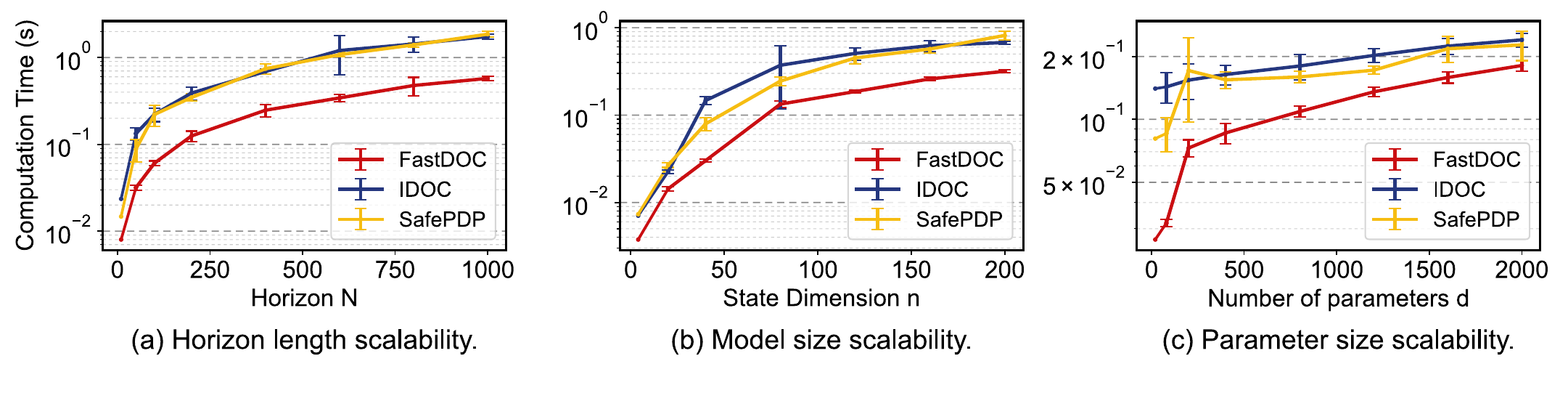}
		\caption{Scalability of \texttt{FastDOC}, \texttt{IDOC}, and \texttt{SafePDP} on synthetic benchmarks. (a) Runtime versus horizon length $N$ (up to $1000$). 
			(b) Runtime versus state dimension $n$ (up to $200$). 
			(c) Runtime versus parameter size $d$ (up to $2000$).}
		\label{fig:Numerical_result}
	\end{center}
\end{figure*}

As illustrated in Fig.~\ref{fig:Numerical_result}, \texttt{IDOC} and \texttt{SafePDP} exhibit nearly identical computation times for solving the full derivatives, while our method consistently outperforms both baselines across all tested dimensions.
The numerical discrepancy between our method and the baseline solvers is negligible, with the relative error remaining below \(10^{-6}\) across all test cases.
In total, our method achieves an average speedup of 2.80 times over \texttt{IDOC} and 2.35 times over \texttt{SafePDP}.

\section{Application to Imitation Learning for Human-Like Autonomous Driving Trajectory Tracking Control}
\label{sec:human-like}
Finally, we demonstrate the practical utility of \texttt{FastDOC} in an NMPC-based imitation learning task for autonomous driving, where the controller learns to imitate human driving behavior and achieve human-like trajectory tracking.

\subsection{NMPC-Based Imitation Learning via FastDOC}
\label{sec:E2E}
\vspace{-0.6em}
\begin{figure}[h]
	\begin{center}
		\includegraphics[width=\linewidth]{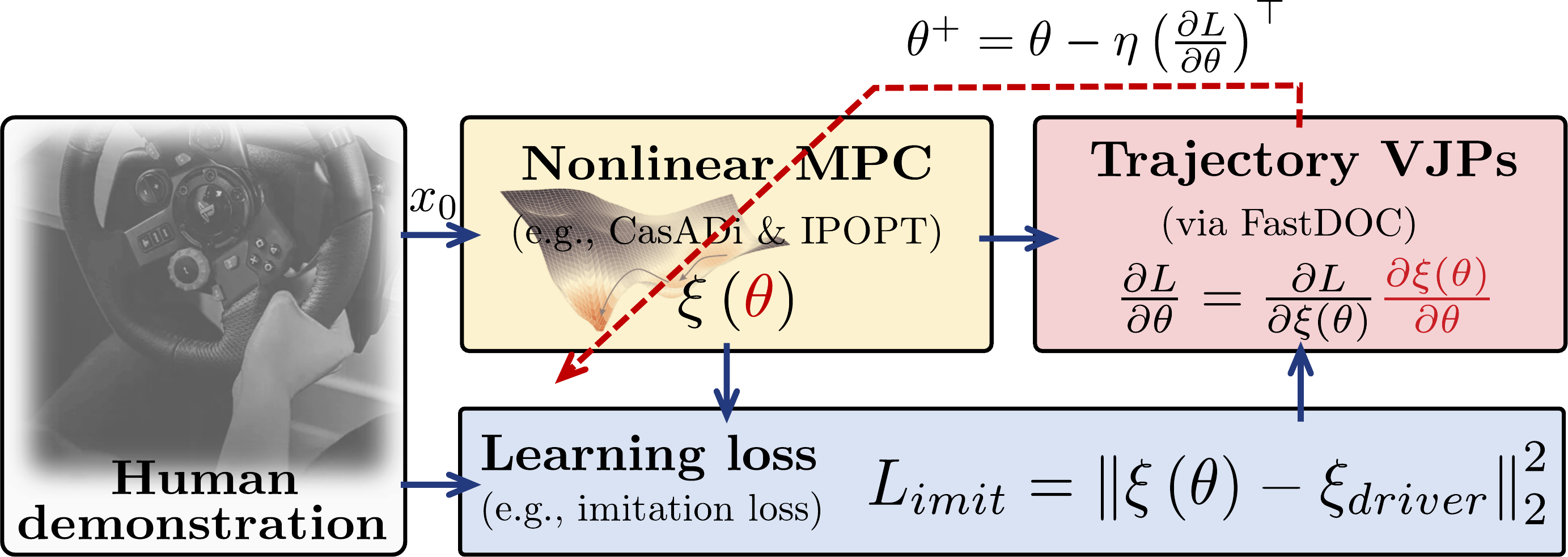}
		\caption{Framework of Imitation Learning through Differentiable NMPC Policy}
		\label{fig:ilframework}
	\end{center}
\end{figure}

Given a human demonstration trajectory $\xi_{\mathrm{driver}}$, the NMPC controller parameterized by $\theta$ generates a nominal trajectory $\xi(\theta)$ under the same initial conditions. In the imitation learning setting, the goal is to adjust $\theta$ such that the controller reproduces the demonstrated behavior. To this end, we minimize a trajectory-level loss that measures the discrepancy between $\xi(\theta)$ and $\xi_{\mathrm{driver}}$ as follows
\begin{equation}\label{eq:loss}
	L = \left\| \xi(\theta) - \xi_{driver} \right\|_2^2.
\end{equation}
Since $\xi(\theta)$ depends on $\theta$, its gradient follows:
\begin{equation}\label{eq:grad_chain}
	\frac{\partial L}{\partial \theta}
	= \frac{\partial L}{\partial \xi(\theta)}
	\frac{\partial \xi(\theta)}{\partial \theta}.
\end{equation}
This gradient provides the update direction for $\theta$ to behave performance like human driver, and the corresponding gradient descent update step:
\begin{equation}
	\theta^{+}=\theta-\eta \left( \frac{\partial L}{\partial \theta} \right).
\end{equation}
The complete procedure is shown in Fig.~\ref{fig:ilframework}.


\subsection{Data Acquisition}

\begin{figure}[h]
	\begin{center}
		\hspace{0.4cm}
		\includegraphics[width=0.9\linewidth, trim=0cm 0cm 0.5cm 0cm, clip]{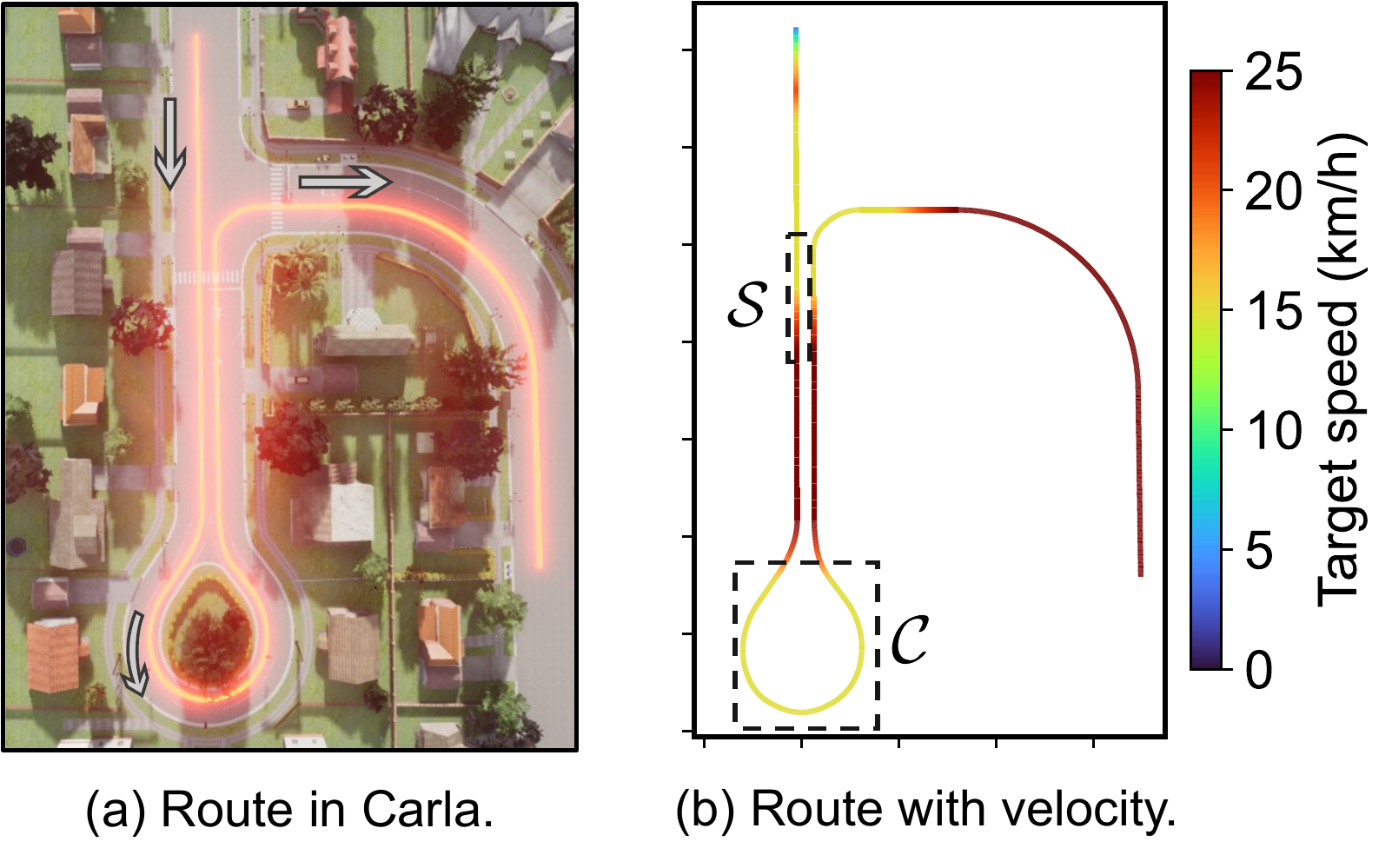}
		\caption{Data acquisition in Carla simulator: a) the route containing straight section $\mathcal{S}$ and curve section $\mathcal{C}$. b) the color-coded target velocity along the route.}
		\label{fig:data-ac}
	\end{center}
\end{figure}
\begin{figure}
	\begin{center}
		\includegraphics[width=\linewidth, trim=12cm 12cm 12cm 12cm, clip]{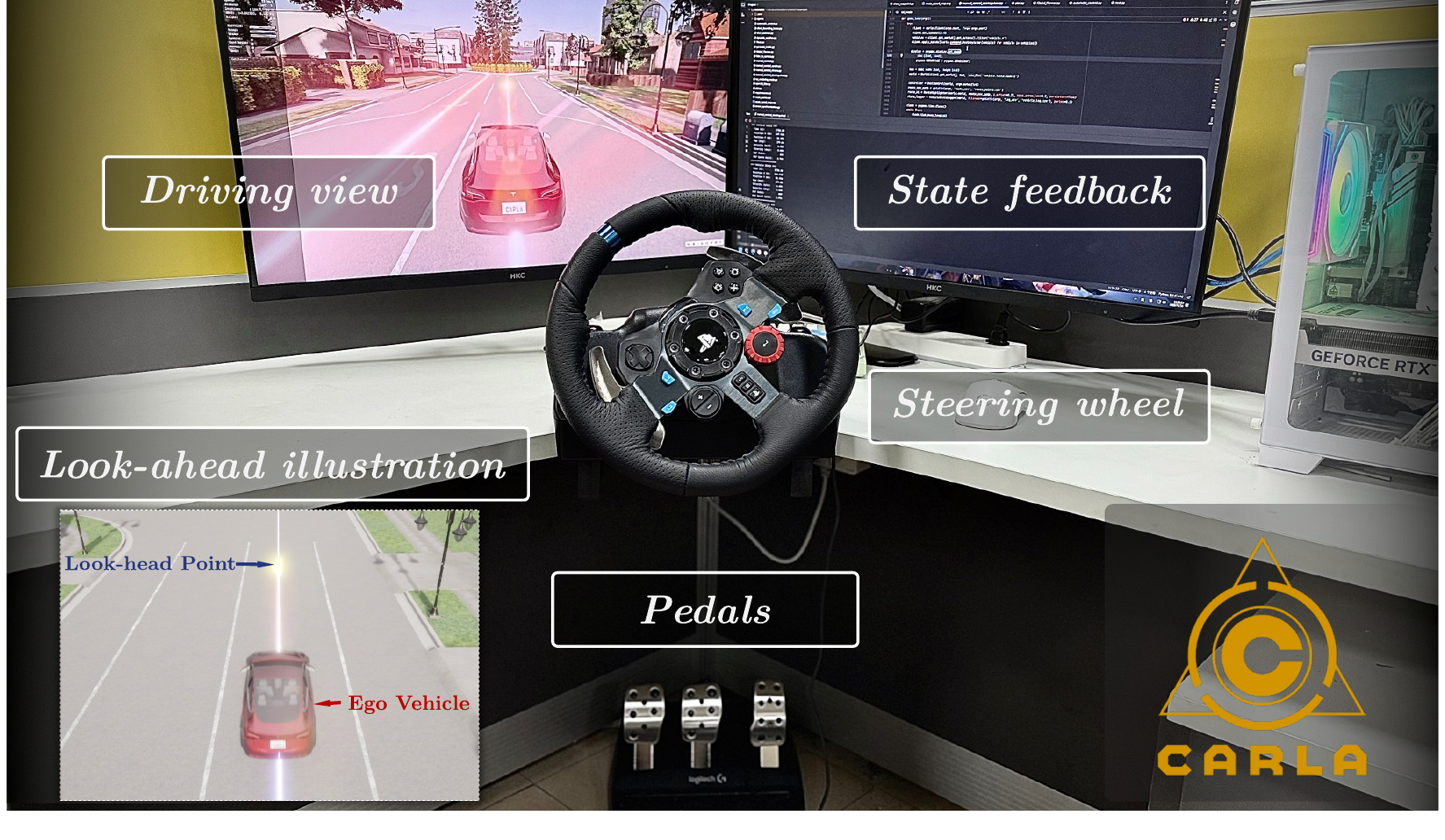}
		\caption{Human-in-the-Loop driving simulator.}
		\label{fig:hardware}
	\end{center}
\end{figure}

This section describes the collection of human driving demonstrations used for imitation learning. As illustrated in Fig.~\ref{fig:data-ac}, the driving scenario constructed in the CARLA environment consists of a straight segment $\mathcal{S}$ with varying target speeds and a high-curvature roundabout $\mathcal{C}$. Participants were instructed to perform a trajectory-following task and to drive in a manner consistent with their natural driving style. In segment $\mathcal{S}$, the target speed increases to 25~km/h, whereas in $\mathcal{C}$ the target speed is set to 15~km/h. During transitions between speed levels, the reference longitudinal acceleration is limited to 1.2~m/s\(^2\). All demonstration data were collected using the hardware platform depicted in Fig.~\ref{fig:hardware}, which consists of a Logitech G29 force-feedback steering wheel with pedals, a display for rendering the driving view, and a virtual road environment generated by the CARLA simulator. During this session, CARLA provides real-time state feedback, while control commands and road information are exchanged in a closed-loop setting with human drivers in the loop.

\subsection{Formulation of A Parametric Optimal Trajectory Tracking Control Policy}
To enable end-to-end imitation learning of human-like trajectory tracking, we model the controller as a parametric and differentiable policy defined by the following nonlinear optimal control problem (OCP):
\begin{equation}
	\label{eq:trackingOCP}
	\begin{aligned}
		\mathop{\text{minimize}}_{x,u}\quad 
		& \sum_{k=0}^{N-1} \ell_k(x_k,u_k) + \ell_N(x_N) \\
		\text{subject\ to}\quad
		& x_{0} = x_{\mathrm{init}}, \\
		& x_{k+1} = f_k(x_k, u_k), \quad k \in \mathbb{N}_{[0,N-1]}, \\
		& \underline{x} \le x_k \le \overline{x}, \quad k \in \mathbb{N}_{[0,N]}, \\
		& \underline{u} \le u_k \le \overline{u}, \quad k \in \mathbb{N}_{[0,N-1]},
	\end{aligned}
\end{equation}
where the state and control vectors are defined as
$
x_k =
\begin{bmatrix}
	X_k & Y_k & \varphi_k & v_k & \dot v_k & \dot\delta_k
\end{bmatrix}^{\!\top},
u_k =
\begin{bmatrix}
	\ddot v_k & \ddot\delta_k
\end{bmatrix}^{\!\top},
$
where \( (X_k, Y_k) \) denotes the vehicle position, \( \varphi_k \) the yaw angle,  
\( v_k \) the longitudinal velocity, and \( \delta_k \) the steering angle.
The system dynamics $f_k(x_k,u_k)$ are given by
\begin{equation}
	\label{eq:vehicle_model_compact}
	\resizebox{0.88\hsize}{!}{$
	x_{k+1}
= x_k
+ \Delta t\,
\begin{bmatrix}
	v_k \cos\varphi_k,
	v_k \sin\varphi_k,
	\frac{v_k}{L_d}\tan\delta_k,
	\dot v_k,
	\ddot v_k,
	\dot\delta_k
\end{bmatrix}^{\!\top}.$}
\end{equation}

For the design of the stage cost \(\ell_k\), we incorporate key aspects of human driving behavior, 
including look-ahead perception, tracking accuracy, and driving comfort. 
To model the driver’s look-ahead behavior, we define the look-ahead point as
\begin{equation}
\begin{bmatrix}
	X_{\mathrm{la},k} \\
	Y_{\mathrm{la},k}
\end{bmatrix}
=
\begin{bmatrix}
	X_k + D \cos(\varphi_k + \alpha \delta_k) \\
	Y_k + D \sin(\varphi_k + \alpha \delta_k)
\end{bmatrix},
\end{equation}
where \(D\) is the look-ahead distance and \(\alpha\) captures the coupling between steering and the look-ahead orientation.

To jointly encode tracking accuracy and comfort-related quantities in the stage cost, we define the feature vector
$
\tau_k =
\begin{bmatrix}
	X_{\mathrm{la},k} &
	Y_{\mathrm{la},k} &
	\varphi_k &
	v_k &
	\delta_k &
	\dot v_k &
	\ddot v_k &
	\dot\delta_k
\end{bmatrix}^{\!\top},
$
and formulate the stage cost as a weighted quadratic penalty,
\begin{equation}
\ell_k = \| \tau_k - \tau_{\mathrm{ref},k} \|_{W}^2,
\qquad
W = \mathrm{diag}(w),
\end{equation}
where \(\tau_{\mathrm{ref},k}\) denotes the reference feature vector constructed from the prescribed reference trajectory (e.g., the road centerline), and \(w\) contains the personalized weights reflecting the driver’s sensitivity to different aspects of tracking and comfort. We collect all learnable parameters into
$
\theta = (w,\, D,\, \alpha),
$
which together characterize individual driving habits.

With this formulation, the parametric nonlinear OCP in \eqref{eq:trackingOCP} 
serves as a differentiable and learnable policy for end-to-end imitation learning of human demonstrations.

\subsection{Experimental Results and Discussion}

The main hyperparameters, personalization settings, and physical constraints used in the NMPC-based imitation learning experiments are summarized in Table~\ref{tab:setup}. In particular, the prediction horizon is fixed to $N_p = 5$ for both straight and cornering scenarios. The table further lists the training setup (maximum number of iterations and learning rate), the initial personalization parameters, and the admissible ranges of the vehicle states and inputs.

\begin{table}[h]
	\centering
	\caption{Experimental parameters setting.}
	\label{tab:setup}
	\begin{tabular}{lcc}
		\hline
		\textbf{Parameter} & Straight & Curve \\
		\hline
		\textbf{Training setup} & & \\
		Max iterations & 1000 & 300 \\
		Learning rate & 0.01 & 0.001 \\
		\hline
		\textbf{Personalized model} & \multicolumn{2}{c}{} \\
		$Q_{\text{init}}$ & \multicolumn{2}{c}{$\text{diag}([16, 16, 4, 8, 2, 1, 0.5, 1])$} \\
		$D_{\text{init}}$ & \multicolumn{2}{c}{14} \\
		$\alpha_{\text{init}}$ & \multicolumn{2}{c}{1} \\
		\hline
		\textbf{Physical constraints} & \multicolumn{2}{c}{} \\
		Speed $v$ & \multicolumn{2}{c}{$0 \leq v \leq 30$ (km/h)} \\
		Acceleration $a$ & \multicolumn{2}{c}{$-2 \leq a \leq 2$ (m/s$^2$)} \\
		Jerk $\dot{a}$ & \multicolumn{2}{c}{$-4 \leq \dot{a} \leq 4$ (m/s$^3$)} \\
		Steering $\delta$ & \multicolumn{2}{c}{$-\pi/6 \leq \delta \leq \pi/6$ (rad)} \\
		Steering rate $\dot{\delta}$ & \multicolumn{2}{c}{$-1.5 \leq \dot{\delta} \leq 1.5$ (rad/s)} \\
		\hline
	\end{tabular}
\end{table}

\begin{figure}[h]
	\begin{center}
		\includegraphics[width=\linewidth]{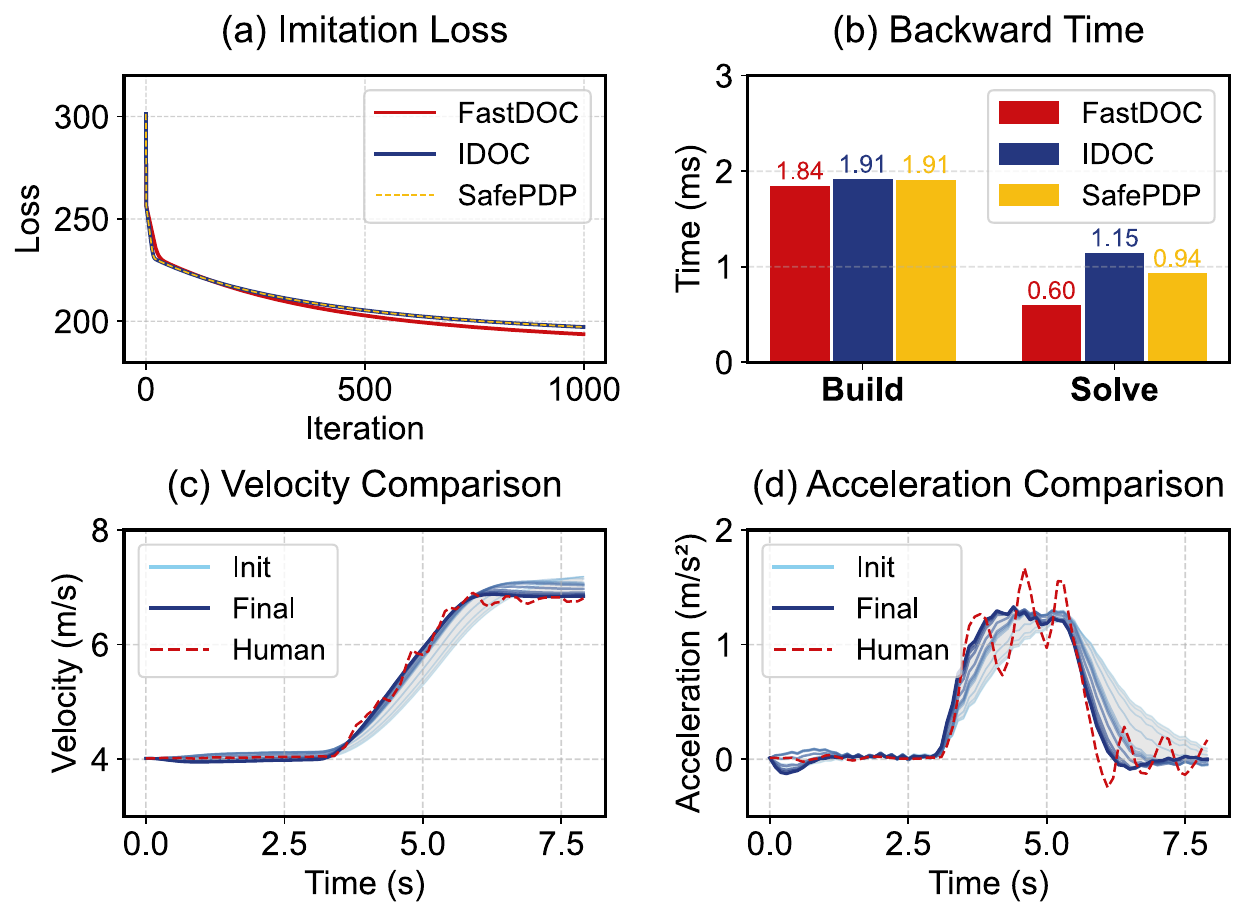}
		\caption{Results of imitation learning under straight road section.}
		\label{fig:straight_result}
	\end{center}
\end{figure}

In the straight road section~$\mathcal{S}$, Fig.~\ref{fig:straight_result}(a) shows that the loss decreases steadily and converges to a value comparable to the baselines. This indicates
that the Gauss–Newton approximation adopted in
\texttt{FastDOC} is not overly restrictive. It should note that because the demonstrations are not strictly realizable (optimal) under the parameterized objective function, the final loss for the \texttt{FastDOC} is small but not zero, which explains that given sub-optimal demonstrations, \texttt{FastDOC} can still find the ‘best’ control objective function within the function set $J(\theta)$ such that its reproduced $\xi({\theta})$ has the minimal distance to the demonstrations.
In Fig.~\ref{fig:straight_result}(b), \textbf{Build} refers to constructing the matrices 
$H$, $A$, $B$, $C$ in the explicit formulation~\eqref{eq:explicit}, whereas \textbf{Solve} denotes computing the trajectory derivatives 
$\nabla_{\theta} \xi$ by solving the corresponding linear systems.
Because \texttt{FastDOC} avoids analytical Hessian computation during the building stage and exploits matrix structure during the solving stage, the overall backward computation time is significantly reduced.
This directly accelerates the NMPC-based imitation learning process.
Regarding imitation performance, Fig.~\ref{fig:straight_result}(c)--(d) show the evolution of the velocity and acceleration trajectories during training. 
By learning from human demonstrations, the NMPC controller does not merely drive the state trajectory closer to the human data; instead, it adapts the underlying parameter set~$\theta$ that shapes the cost function, thereby capturing the driver’s style. 
Through this parameter adaptation, the resulting trajectories exhibit human-like tracking behavior in the straight section.

\begin{figure}[h]
	\begin{center}
		\includegraphics[width=\linewidth]{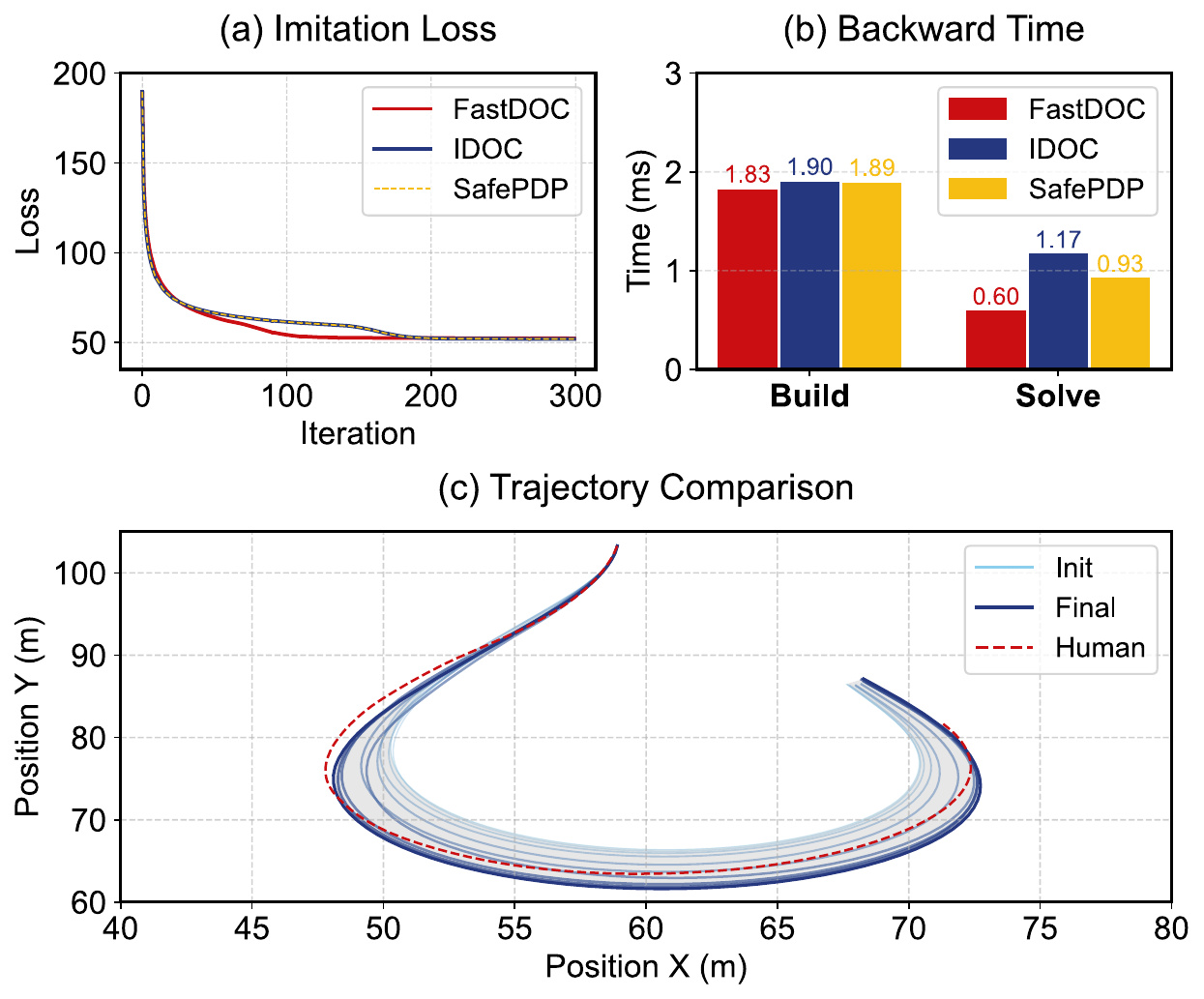}
		\caption{Results of Imitation Learning under curved road section.}
		\label{fig:curve_result}
	\end{center}
\end{figure}
In the curve road section $\mathcal{C}$, our NMPC-based imitation learning result is shown in Fig.~\ref{fig:curve_result}. 
The curve scenario introduces stronger nonlinearities compared with the straight section, yet the experimental outcomes remain consistent with those observed previously. 
(a) The imitation loss decreases and converges to a value comparable to the baselines. 
(b) \texttt{FastDOC} achieves lower computation time in both the building and solving stages. 
(c) The generated trajectories from NMPC after training align more closely with the human driving demonstrations. 
These results indicate that \texttt{FastDOC} can still provide reliable descent directions under stronger nonlinearities, enabling the controller to learn personalized parameters and exhibit human-like tracking behavior in the curve scenario.

Across both scenarios, the experiments evaluate two key considerations for practical deployment: 
the reliability of the Gauss-Newton approximation under small residuals and the solver's stability in highly nonlinear regimes. 
The results show that \texttt{FastDOC} handles both effectively, enabling the controller to adapt toward human driving behavior and demonstrating its practical usefulness.

\section{Conclusion}
We propose nonlinear \texttt{FastDOC}, an efficient method for computing trajectory derivatives of nonlinear OCP. Under a Gauss–Newton approximation, \texttt{FastDOC} fully exploits the special structure of the matrices that arise in the analytical formulation of trajectory derivatives, leading to substantial improvements in computational efficiency by accelerating the most time-consuming matrix factorizations. Numerical results on comprehensive benchmarks demonstrate that \texttt{FastDOC} achieves an average speedup of 1.8 times compared to the baseline. Additionally, we evaluate our method on an NMPC-based imitation-learning task in autonomous driving, where the results further validate the practical effectiveness of \texttt{FastDOC}. 

\bibliography{ifacconf}             
\end{document}